# Future Solar Neutrino Experiments and Neutrino Spin-Flavour Precession

S.M. Bilenky[(a,b,c)*] and C. Giunti[(b,c)⋆]

[(a)] Joint Institute of Nuclear Research, Dubna, Russia
[(b)] INFN Torino, Via P. Giuria 1, I–10125 Torino, Italy
[(c)] Dipartimento di Fisica Teorica, Università di Torino

Abstract

The main features of the observables in the SNO and Super-Kamiokande solar neutrino experiments in the case of neutrino spin and/or spin-flavour precession in the magnetic field of the sun are discussed. It is shown without any model dependent assumption that in the case of Majorana transition magnetic moments the NC event rate $N^{\rm NC}$ does not depend on time and a measurement of $N^{\rm NC}$ will allow to determine the initial flux of $^8\mathrm{B}$ neutrinos with a theoretical uncertainty of a few %. In the case of Dirac magnetic moments the NC event rate will depend on time and we obtain a model independent lower bound for the transition probability of initial $\nu_e$'s into right-handed sterile neutrinos.

* BILENKY@TO.INFN.IT
⋆ GIUNTI@TO.INFN.IT

# 1 Introduction

It was shown in ref.[1, 2, 3] that future real-time solar neutrino experiments (SNO [4] and others), in which high-energy neutrinos from $^8\mathrm{B}$ decay will be detected through the observation of **different** reactions, will allow to separate the investigation of neutrino properties (masses, mixing, etc.) from the investigation of the central invisible region of the sun in which energy is generated.

If only active neutrinos are present in the solar neutrino flux on the earth, the initial flux of $^8\mathrm{B}$ neutrinos and the $\nu_e$ survival probability can be determined directly from the data of future solar neutrino experiments without any assumption about the mechanism of transition of $\nu_e$ into $\nu_\mu$ and/or $\nu_\tau$. In the general case of neutrino mixing, solar $\nu_e$'s can transfer into active neutrinos $\nu_e$, $\nu_\mu$, $\nu_\tau$ and into sterile left-handed (anti)neutrinos (see ref.[5]). Different model independent relations and inequalities between observables, which could allow to reveal whether there are sterile neutrinos in the flux of solar neutrinos on the earth, were derived in ref.[3].

In this paper we will discuss possibilities of a model independent treatment of the data from future solar neutrino experiments in the case of spin and/or spin-flavour precession of solar $\nu_e$'s due to anomalously large neutrino magnetic moments. The effects of neutrino magnetic moments [6, 7] were widely discussed in last years (see ref.[8]) in connection with a possible indication of the existence of an anticorrelation between the flux of solar $\nu_e$'s and the sunspot number [9].

The possible transitions of neutrinos in the magnetic field of the sun depend on the nature of neutrinos. In the case of Dirac neutrinos, solar $\nu_e$'s can be transformed into sterile right-handed neutrinos $\nu_{\ell R}$ ($\ell = e, \mu, \tau$), quanta of right-handed fields. In the case of Majorana neutrinos, transitions of solar $\nu_e$'s into active right-handed antineutrinos $\bar{\nu}_\ell$ ($\ell = e, \mu, \tau$), quanta of left-handed fields, could take place. Notice that direct transitions $\nu_e \to \bar{\nu}_e$ are forbidden by CPT invariance. However, sizable $\nu_e \to \bar{\nu}_e$ transitions can occur under special conditions if both spin-flavour precession and the MSW or vacuum oscillations mechanisms are operating [10][1].

In the SNO experiment neutrinos (antineutrinos) from the sun will be detected through observation of the following processes:

$$\nu_e + d \to e^- + p + p \qquad (\mathrm{CC}) \tag{1}$$

$$\nu\,(\bar{\nu}) + d \to \nu\,(\bar{\nu}) + p + n \qquad (\mathrm{NC}) \tag{2}$$

$$\nu\,(\bar{\nu}) + e^- \to \nu\,(\bar{\nu}) + e^- \qquad (\mathrm{ES}) \tag{3}$$

It is also planned [13] to search for $\bar{\nu}_e$ from the sun with the help of the reaction

$$\bar{\nu}_e + d \to e^+ + n + n \tag{4}$$

The threshold for neutrino detection in this experiment will be rather high ($E_{\mathrm{th}} \gtrsim 5\,\mathrm{MeV}$ for CC and ES, $E_{\mathrm{th}} = 2.2\,\mathrm{MeV}$ for NC and $E_{\mathrm{th}} \geq 4\,\mathrm{MeV}$ for reaction (4)). Thus SNO will detect $^8\mathrm{B}$

[1] Let us notice that the Mont Blanc collaboration has obtained the following upper bound for the flux of $\bar{\nu}_e$ in the energy range $9\,\mathrm{MeV} \leq E \leq 20\,\mathrm{MeV}$: $\Phi_{\bar{\nu}_e} < 8.2 \times 10^4\,\mathrm{cm}^{-2}\,\mathrm{sec}^{-1}$ [11]. From the analysis of the background in the Kamiokande experiment the following upper bound for the flux of high-energy $\bar{\nu}_e$ with $E \geq 10.6\,\mathrm{MeV}$ was obtained: $\Phi_{\bar{\nu}_e} < 6.1 \times 10^4\,\mathrm{cm}^{-2}\,\mathrm{sec}^{-1}$ [12].

$\nu_e$'s and neutrinos (and antineutrinos) originating from them. The energy spectrum of the initial $^8\mathrm{B}$ $\nu_e$'s is given by

$$\phi_{\nu_e}^{^8\mathrm{B}}(E) = \Phi_{\nu_e}^{^8\mathrm{B}} X(E) \tag{5}$$

where $X(E)$ is a known function (the phase space factor of the decay $^8\mathrm{B} \to {}^8\mathrm{Be} + e^+ + \nu_e$) and $\Phi_{\nu_e}^{^8\mathrm{B}}$ is the total initial flux (determined by the central temperature of the sun, the cross sections of different reactions of the $pp$ and CNO cycles, etc.). In the Super-Kamiokande (S-K) experiment [14] high energy solar neutrinos ($E_{\mathrm{th}} \gtrsim 5\,\mathrm{MeV}$) will be detected through the observation of process (3).

A measurement of the CC event rate as a function of neutrino energy $E$ will allow to obtain the $\nu_e$ flux on the earth $\phi_{\nu_e}(E)$ and to determine the $\nu_e$ survival probability up to a constant. If it will occur that the CC event rate depends on time periodically, it will be an evidence that neutrinos have large magnetic moments and their effects are important. In this paper we discuss which additional informations can be extracted from a measurement of the NC and ES event rates through the observation of the processes (2) and (3), respectively. In particular, we will show that through the observation of the NC and ES reactions it will be possible to distinguish Dirac from Majorana magnetic moments.

# 2 Neutral Current

Let us consider first the NC process (2). In the case of Majorana neutrino magnetic moments the integral NC event rate is given by

$$N^{\mathrm{NC}} = \int_{E_{\mathrm{th}}} \sigma_{\nu d}^{\mathrm{NC}}(E) \sum_{\ell=e,\mu,\tau} \phi_{\nu_\ell}(E)\,\mathrm{d}E + \int_{E_{\mathrm{th}}} \sigma_{\bar{\nu} d}^{\mathrm{NC}}(E) \sum_{\ell=e,\mu,\tau} \phi_{\bar{\nu}_\ell}(E)\,\mathrm{d}E \tag{6}$$

where $\sigma_{\nu d}^{\mathrm{NC}}(E)$ and $\sigma_{\bar{\nu} d}^{\mathrm{NC}}(E)$ are the cross sections of the processes $\nu d \to \nu p n$ and $\bar{\nu} d \to \bar{\nu} p n$, $\sum_{\ell=e,\mu,\tau} \phi_{\nu_\ell}(E)$ and $\sum_{\ell=e,\mu,\tau} \phi_{\bar{\nu}_\ell}(E)$ are the fluxes of all types of neutrinos and antineutrinos on the earth. Taking into account that

$$\sum_{\ell=e,\mu,\tau} \phi_{\nu_\ell}(E) + \sum_{\ell=e,\mu,\tau} \phi_{\bar{\nu}_\ell}(E) = \phi_{\nu_e}^{^8\mathrm{B}}(E) \tag{7}$$

from Eq.(5) and Eq.(6) we obtain

$$\frac{N^{\mathrm{NC}}}{\Phi_{\nu_e}^{^8\mathrm{B}}\,\overline{X}^{\mathrm{NC}}} = 1 + I^{\mathrm{NC}} \tag{8}$$

where

$$I^{\mathrm{NC}} \equiv \frac{1}{\overline{X}^{\mathrm{NC}}} \int_{E_{\mathrm{th}}} \left[ \frac{\sigma_{\nu d}^{\mathrm{NC}}(E) - \sigma_{\bar{\nu} d}^{\mathrm{NC}}(E)}{2} \right] X(E) \sum_{\ell=e,\mu,\tau} \left[ \mathrm{P}_{\nu_e \to \nu_\ell}(E) - \mathrm{P}_{\nu_e \to \bar{\nu}_\ell}(E) \right] \mathrm{d}E \tag{9}$$

Here $\sum_{\ell=e,\mu,\tau} \mathrm{P}_{\nu_e\to\nu_\ell}(E) \left( \sum_{\ell=e,\mu,\tau} \mathrm{P}_{\nu_e\to\bar{\nu}_\ell}(E) \right)$ is the transition probability of solar $\nu_e$'s into neutrinos (antineutrinos) of all types and

$$\overline{X}^{\mathrm{NC}} \equiv \frac{X_{\nu d}^{\mathrm{NC}} + X_{\bar{\nu} d}^{\mathrm{NC}}}{2} \tag{10}$$

$$X_{\nu d(\bar{\nu} d)}^{\mathrm{NC}} \equiv \int_{E_{\mathrm{th}}} \sigma_{\nu d(\bar{\nu} d)}^{\mathrm{NC}}(E)\, X(E)\, \mathrm{d}E \tag{11}$$

The cross sections of the processes $\nu d \to \nu n p$ and $\bar{\nu} d \to \bar{\nu} n p$ where calculated by several groups and reviewed in ref.[15]. Using the results presented in ref.[15], we obtained [2]

$$X_{\nu d}^{\mathrm{NC}} = 4.72 \times 10^{-43}\,\mathrm{cm}^2 \tag{12}$$

$$X_{\bar{\nu} d}^{\mathrm{NC}} = 4.51 \times 10^{-43}\,\mathrm{cm}^2 \tag{13}$$

It is easy to see that the value of the integral $I^{\mathrm{NC}}$ is very small. In fact, for the absolute value of this integral we have the following upper bound:

$$\left| I^{\mathrm{NC}} \right| \leq \frac{X_{\nu d}^{\mathrm{NC}} - X_{\bar{\nu} d}^{\mathrm{NC}}}{X_{\nu d}^{\mathrm{NC}} + X_{\bar{\nu} d}^{\mathrm{NC}}} \simeq 2 \times 10^{-2} \tag{14}$$

The upper bound of the integral $I^{\mathrm{NC}}$ is so small because the cross sections of the processes $\nu d \to \nu n p$ and $\bar{\nu} d \to \bar{\nu} n p$ are very close to each other in the energy region near the threshold. The argument in favour of this fact is rather general: from symmetry considerations it follows that near the threshold (if only the s-state of the final nucleons is taken into account) the vector current does not contribute to the matrix elements of the processes $\nu d \to \nu n p$ and $\bar{\nu} d \to \bar{\nu} n p$ and the cross sections $\sigma_{\nu d}^{\mathrm{NC}}$ and $\sigma_{\bar{\nu} d}^{\mathrm{NC}}$ are equal. The corrections due to higher states are small in the relevant energy region (see ref.[15] and references therein). Thus, the term $I^{\mathrm{NC}}$ in Eq.(8) can be safely neglected and, in the case of Majorana magnetic moments, we come to the following conclusions:

1. The NC event rate does not depend on time (within less than 2%).

2. The flux of the initial $^8\mathrm{B}$ $\nu_e$'s is given by

$$\Phi_{\nu_e}^{^8\mathrm{B}} \simeq \frac{N^{\mathrm{NC}}}{\overline{X}^{\mathrm{NC}}} \tag{15}$$

   and therefore can be determined directly from the experimental data[3].

[2] We used the values of the function $X(E)$ given in ref.[16].

[3] Notice that the expression (15) for $\Phi_{\nu_e}^{^8\mathrm{B}}$ coincides with that obtained in ref.[2] for the case of transitions of solar $\nu_e$'s only into $\nu_\mu$ and/or $\nu_\tau$ due to usual neutrino mixing.

3. It is possible to obtain the $\nu_e$ survival probability directly from measurable quantities:

$$\begin{aligned} \mathrm{P}_{\nu_e\to\nu_e}(E) &= 1 - \sum_{\ell=\mu,\tau} \mathrm{P}_{\nu_e\to\nu_\ell}(E) - \sum_{\ell=e,\mu,\tau} \mathrm{P}_{\nu_e\to\bar\nu_\ell}(E) \\ &= \frac{\phi_{\nu_e}(E)\,\overline{X}^{\mathrm{NC}}}{X(E)\,N^{\mathrm{NC}}} \end{aligned} \tag{16}$$

where $\phi_{\nu_e}(E)$ is the flux of $\nu_e$ on the earth, which can be determined from the CC event rate.

Consider now the case of Dirac neutrino magnetic moments. In this case

$$N^{\mathrm{NC}} = \int_{E_{\mathrm{th}}} \sigma^{\mathrm{NC}}_{\nu d}(E) \sum_{\ell=e,\mu,\tau} \phi_{\nu_\ell}(E)\,\mathrm{d}E \tag{17}$$

where $\sum_{\ell=e,\mu,\tau} \phi_{\nu_\ell}(E) = \phi^{^8\mathrm{B}}_{\nu_e}(E) - \phi_{\nu_S}(E)$ and $\phi_{\nu_S}(E)$ is the flux of sterile right-handed neutrinos on the earth. It is clear that in the Dirac case $N^{\mathrm{NC}}$ will depend on time. Hence, a time dependence of the CC **and** NC event rates will be a signal that neutrinos have large Dirac magnetic moments.

For the average transition probability of solar $\nu_e$'s into sterile states we obtain the following model independent lower bound:

$$\begin{aligned} \left\langle \sum_{\ell=e,\mu,\tau} \mathrm{P}_{\nu_e\to\nu_{\ell R}}(E) \right\rangle_{\mathrm{NC}} &\equiv \frac{1}{X^{\mathrm{NC}}_{\nu d}} \int_{E_{\mathrm{th}}} \sigma^{\mathrm{NC}}_{\nu d}(E)\,X(E) \sum_{\ell=e,\mu,\tau} \mathrm{P}_{\nu_e\to\nu_{\ell R}}(E)\,\mathrm{d}E \\ &\geq 1 - \frac{N^{\mathrm{NC}}}{X^{\mathrm{NC}}_{\nu d}\ (\phi_{\nu_e}/X)_{\mathrm{max}}} \end{aligned} \tag{18}$$

where $(\phi_{\nu_e}/X)_{\mathrm{max}}$ indicates the maximum value of $\phi_{\nu_e}(E)/X(E)$, which can be obtained from the data on the CC event rate. Let us stress that, unlike the case considered in ref.[3], in the case of spin and/or spin-flavour transitions due to Dirac magnetic moments the lower bound (18) will depend on time.

# 3 Elastic Scattering

Let us consider now the ES process (3). Using Eq.(5), in the case of Majorana neutrinos we have

$$\frac{\widetilde{\Sigma}^{\mathrm{ES}}}{\Phi^{^8\mathrm{B}}_{\nu_e}\,\overline{X}^{\mathrm{ES}}} = 1 + I^{\mathrm{ES}} \tag{19}$$

where

$$\begin{aligned} \widetilde{\Sigma}^{\mathrm{ES}} \equiv\ N^{\mathrm{ES}} &- \int_{E_{\mathrm{th}}} \Big[\sigma_{\nu_e e}(E) - \sigma_{\nu_\mu e}(E)\Big]\,\phi_{\nu_e}(E)\,\mathrm{d}E \\ &- \int_{E_{\mathrm{th}}} \Big[\sigma_{\bar\nu_e e}(E) - \sigma_{\bar\nu_\mu e}(E)\Big]\,\phi_{\bar\nu_e}(E)\,\mathrm{d}E\ , \end{aligned} \tag{20}$$

$N^{\rm ES}$ is the integral ES event rate, $\sigma_{\nu_\ell e}(E)$ ($\sigma_{\bar\nu_\ell e}(E)$) is the total cross section of the process $\nu_\ell e \to \nu_\ell e$ ($\bar\nu_\ell e \to \bar\nu_\ell e$) with $\ell = e, \mu$, and

$$I^{\rm ES} \equiv \frac{1}{\overline{X}^{\rm ES}} \int_{E_{\rm th}} \left[ \frac{\sigma_{\nu_\mu e}(E) - \sigma_{\bar\nu_\mu e}(E)}{2} \right] X(E) \sum_{\ell=e,\mu,\tau} \left[ {\rm P}_{\nu_e\to\nu_\ell}(E) - {\rm P}_{\nu_e\to\bar\nu_\ell}(E) \right] {\rm d}E \qquad (21)$$

where

$$\overline{X}^{\rm ES} \equiv \frac{X^{\rm ES}_{\nu_\mu e} + X^{\rm ES}_{\bar\nu_\mu e}}{2} \qquad (22)$$

$$X^{\rm ES}_{\nu_\mu e(\bar\nu_\mu e)} \equiv \int_{E_{\rm th}} \sigma_{\nu_\mu e(\bar\nu_\mu e)}(E)\, X(E)\, {\rm d}E \qquad (23)$$

The values of the cross sections $\sigma_{\nu_\ell e}(E)$ and $\sigma_{\bar\nu_\ell e}(E)$ ($\ell = e, \mu$) with a 5 MeV threshold kinetic energy for electron detection are depicted in Fig.1 in the energy range relevant for solar neutrino experiments. We used $g_V = -0.036$ and $g_A = -0.505$ [17]. For $X^{\rm ES}_{\nu_\mu e}$ and $X^{\rm ES}_{\bar\nu_\mu e}$ we obtained the following values:

$$X^{\rm ES}_{\nu_\mu e} = 2.71 \times 10^{-45}\,{\rm cm}^2 \qquad (24)$$

$$X^{\rm ES}_{\bar\nu_\mu e} = 2.14 \times 10^{-45}\,{\rm cm}^2 \qquad (25)$$

Using these values, we have the following upper bound for the absolute value of the integral $I^{\rm ES}$:

$$\left| I^{\rm ES} \right| \leq \frac{X^{\rm ES}_{\nu_\mu e} - X^{\rm ES}_{\bar\nu_\mu e}}{X^{\rm ES}_{\nu_\mu e} + X^{\rm ES}_{\bar\nu_\mu e}} \simeq 0.12 \qquad (26)$$

Let us notice that the real value of $\left| I^{\rm ES} \right|$ can be significantly smaller than the upper bound given in Eq.(26). We calculated the integral $I^{\rm ES}$ in the simplest model with two non-mixed massive Majorana neutrinos and a large transition magnetic moment $\mu_{e\mu}$. In ref.[18] it was shown that the existing solar neutrino data can be described by this model with $\mu_{e\mu} \simeq 10^{-11}\,\mu_{\rm B}$ under specific assumptions for the magnetic field of the sun. For the average value of $\Delta m^2$ found in ref.[18] ($\Delta m^2 = 10^{-8}\,{\rm eV}^2$) we obtained that $-7.6 \times 10^{-2} \leq I^{\rm ES} \leq 9.3 \times 10^{-3}$, where the lower and upper bounds correspond to high and low solar activity, respectively.

The time dependence of the quantity $\widetilde{\Sigma}^{\rm ES}$ is determined by the integral $I^{\rm ES}$, which is less than $\simeq 10\%$. Neglecting $I^{\rm ES}$ in Eq.(19), we obtain the following approximate expression for the initial flux of $^8{\rm B}$ neutrinos:

$$\Phi^{^8{\rm B}}_{\nu_e} \simeq \frac{\widetilde{\Sigma}^{\rm ES}}{\overline{X}^{\rm ES}} \qquad (27)$$

Thus, in the case of Majorana magnetic moments, the initial flux of $^8{\rm B}$ neutrinos can be determined in two independent ways: from the NC event rate (see Eq.(15)) and from the ES and CC event rates (see Eq.(27)). Therefore, independently from the value of the initial neutrino flux, we have the following approximate relation between measurable quantities:

$$N^{\rm NC} \simeq \frac{\overline{X}^{\rm NC}}{\overline{X}^{\rm ES}}\, \widetilde{\Sigma}^{\rm ES} \qquad (28)$$

This relation is a generalization of an analogous relation that was obtained in ref.[2] for the case in which only active neutrinos $\nu_e$, $\nu_\mu$, $\nu_\tau$ are present in the flux of solar neutrinos on the earth (in that case the relation is exact).

Let us notice that, in the case of Majorana magnetic moments, the ES event rate will depend on time. This time dependence is determined by the time dependence of the CC event rate.

Let us consider now the case of Dirac neutrino magnetic moments. In this case we have

$$\frac{\Sigma^{\rm ES}}{\Phi_{\nu_e}^{^8{\rm B}}\, X_{\nu_\mu e}^{\rm ES}} = 1 - \frac{1}{X_{\nu_\mu e}^{\rm ES}} \int_{E_{\rm th}} \sigma_{\nu_\mu e}(E)\, X(E) \sum_{\ell=e,\mu,\tau} {\rm P}_{\nu_e\to\nu_{\ell R}}(E)\, {\rm d}E \tag{29}$$

where

$$\Sigma^{\rm ES} \equiv N^{\rm ES} - \int_{E_{\rm th}} \Big(\sigma_{\nu_e e}(E) - \sigma_{\nu_\mu e}(E)\Big)\, \phi_{\nu_e}(E)\, {\rm d}E \tag{30}$$

In the case under consideration the quantity $\Sigma^{\rm ES}$ will depend on time. From Eq.(29), we obtain the following lower bound for the average transition probability of solar $\nu_e$'s into all possible right-handed sterile states:

$$\begin{aligned} \left\langle \sum_{\ell=e,\mu,\tau} {\rm P}_{\nu_e\to\nu_{\ell R}}(E) \right\rangle_{\rm ES} &\equiv \frac{1}{X_{\nu_\mu e}^{\rm ES}} \int_{E_{\rm th}} \sigma_{\nu_\mu e}(E)\, X(E) \sum_{\ell=e,\mu,\tau} {\rm P}_{\nu_e\to\nu_{\ell R}}(E)\, {\rm d}E \\ &\geq 1 - \frac{\Sigma^{\rm ES}}{X_{\nu_\mu e}^{\rm ES}\, (\phi_{\nu_e}/X)_{\rm max}} \end{aligned} \tag{31}$$

In the case of Dirac magnetic moments, instead of relation (28) we have

$$N^{\rm NC} = \frac{X_{\nu d}^{\rm NC}}{X_{\nu_\mu e}^{\rm ES}}\, \Sigma^{\rm ES} + \beta \tag{32}$$

where

$$\beta = \frac{X_{\nu d}^{\rm NC}}{X_{\nu_\mu e}^{\rm ES}} \int_{E_{\rm th}} \sigma_{\nu_\mu e}(E)\, \phi_{\nu_{\rm S}}(E)\, {\rm d}E - \int_{E_{\rm th}} \sigma_{\nu d}^{\rm NC}(E)\, \phi_{\nu_{\rm S}}(E)\, {\rm d}E \tag{33}$$

Let us assume that $\bar\nu_e$ is not observed. If the relation (28) is violated and $\beta$ depends on time, we will have an additional argument in favour of Dirac magnetic moments. Notice, however, that, according to our model calculations, the two terms in the right-hand side of Eq.(33) could cancel each other.

# 4 On the transition probability of $\nu_e$ into $\bar\nu_\mu$ and/or $\bar\nu_\tau$

Let us consider the case of Majorana magnetic moments (the CC event rate depends on time but the NC event rate is constant). As we have seen in Sec.2, from the data on the CC and NC reactions (and the process (4)) it will be possible to determine the sum of the probabilities of the transitions of initial $\nu_e$'s into $\nu_\mu$, $\nu_\tau$, $\bar{\nu}_\mu$, $\bar{\nu}_\tau$ (see Eq.(16)). In this section we will discuss possibilities to obtain from the SNO and S-K data an information about the transition probability $\sum\limits_{\ell=\mu,\tau} \mathrm{P}_{\nu_e \to \bar{\nu}_\ell}(E)$. With the help of Eq.(15) we obtain

$$\frac{\Omega^{\mathrm{ES}} \, \overline{X}^{\mathrm{NC}}}{N^{\mathrm{NC}} \, X^{\mathrm{ES}}_{\nu_\mu e}} \simeq 1 - \widetilde{I}^{\mathrm{ES}} \tag{34}$$

where

$$\begin{aligned} \Omega^{\mathrm{ES}} \quad \equiv \quad & N^{\mathrm{ES}} - \int_{E_{\mathrm{th}}} \Big[ \sigma_{\nu_e e}(E) - \sigma_{\nu_\mu e}(E) \Big] \, \phi_{\nu_e}(E) \, \mathrm{d}E \\ & - \int_{E_{\mathrm{th}}} \Big[ \sigma_{\bar{\nu}_e e}(E) - \sigma_{\nu_\mu e}(E) \Big] \, \phi_{\bar{\nu}_e}(E) \, \mathrm{d}E \end{aligned} \tag{35}$$

and

$$\widetilde{I}^{\mathrm{ES}} \equiv \frac{1}{X^{\mathrm{ES}}_{\nu_\mu e}} \int_{E_{\mathrm{th}}} \Big[ \sigma_{\nu_\mu e}(E) - \sigma_{\bar{\nu}_\mu e}(E) \Big] \, X(E) \sum_{\ell=\mu,\tau} \mathrm{P}_{\nu_e \to \bar{\nu}_\ell}(E) \, \mathrm{d}E \tag{36}$$

It is easy to see that

$$0 \leq \widetilde{I}^{\mathrm{ES}} \leq \frac{X^{\mathrm{ES}}_{\nu_\mu e} - X^{\mathrm{ES}}_{\bar{\nu}_\mu e}}{X^{\mathrm{ES}}_{\nu_\mu e}} = 0.21 \tag{37}$$

Hence, in order to obtain an information about the transition probability $\sum\limits_{\ell=\mu,\tau} \mathrm{P}_{\nu_e \to \bar{\nu}_\ell}(E)$, it is necessary to know the measurable quantity $\Omega^{\mathrm{ES}} \overline{X}^{\mathrm{NC}} / N^{\mathrm{NC}} X^{\mathrm{ES}}_{\nu_\mu e}$ with an accuracy better than $\simeq 20\%$. If this accuracy will be achieved, the average probability of the transition of solar $\nu_e$'s into $\bar{\nu}_\mu$ and/or $\bar{\nu}_\tau$ can be determined directly from the experimental data:

$$\begin{aligned} \left\langle \sum_{\ell=\mu,\tau} \mathrm{P}_{\nu_e \to \bar{\nu}_\ell}(E) \right\rangle \quad & \equiv \quad \frac{1}{X^{\mathrm{ES}}_{\nu_\mu e} - X^{\mathrm{ES}}_{\bar{\nu}_\mu e}} \int_{E_{\mathrm{th}}} \Big[ \sigma_{\nu_\mu e}(E) - \sigma_{\bar{\nu}_\mu e}(E) \Big] \, X(E) \sum_{\ell=\mu,\tau} \mathrm{P}_{\nu_e \to \bar{\nu}_\ell}(E) \, \mathrm{d}E \\ & \simeq \quad \frac{X^{\mathrm{ES}}_{\nu_\mu e}}{X^{\mathrm{ES}}_{\nu_\mu e} - X^{\mathrm{ES}}_{\bar{\nu}_\mu e}} \left( 1 - \frac{\Omega^{\mathrm{ES}} \, \overline{X}^{\mathrm{NC}}}{N^{\mathrm{NC}} \, X^{\mathrm{ES}}_{\nu_\mu e}} \right) \end{aligned} \tag{38}$$

# 5 Conclusions

In SNO, Super-Kamiokande and other future solar neutrino experiments high energy solar neutrinos will be detected through the observation of CC, NC and ES ($\nu$-$e$ elastic scattering) reactions. If neutrinos have large (Majorana or Dirac) magnetic moments, solar neutrinos can undergo spin and/or spin-flavour precessions in the magnetic field of the sun. In this case the CC event rate will manifest a periodical time dependence.

We have shown that in the case of Majorana neutrino magnetic moments the NC event rate in the SNO experiment ($N^{\mathrm{NC}}$) will not depend on time and a measurement of $N^{\mathrm{NC}}$ will allow to determine in a model independent way the total flux of initial $^8\mathrm{B}$ neutrinos (see Eq.(15)) and the $\nu_e$ survival probability (see Eq.(16)). In the case of Majorana neutrino magnetic moments the ES event rates will depend on time. We have shown, however, that a combination of ES and CC event rates ($\widetilde{\Sigma}^{\mathrm{ES}}$, see Eq.(20)) can have only a minor time dependence. We have also shown that, in the case of Majorana neutrino magnetic moments, there is a relation between the NC, CC and ES event rates (Eq.(28)).

In the case of Dirac neutrino magnetic moments the NC event rate in the SNO experiment, as well as the CC and ES event rates, will depend on time. We have shown that a measurement of the NC (or ES) and CC event rates will allow to determine a (time dependent) lower bound for the average transition probability of solar $\nu_e$'s into all possible right-handed sterile states (see Eq.(18) and Eq.(31)).

Finally, for the case of Majorana magnetic moments we have derived an expression that allows to obtain the average probability of the transition of initial $\nu_e$'s into $\bar{\nu}_\mu$ and/or $\bar{\nu}_\tau$ from measurable quantities (see Eq.(38)). The determination of this average probability will require a measurement of the event rates with rather high accuracy.

## Acknowledgments

It is a pleasure for us to express our gratitude to Wanda Alberico for very useful discussions.

# References

[1] H.H. Chen, Phys. Rev. Lett. 55 (1985) 1534; R.S. Raghavan, S. Pakvasa and B.A. Brown, Phys. Rev. Lett. 57 (1986) 1801; J.N. Bahcall, M. Baldo-Ceolin, D.B. Cline and C. Rubbia, Phys. Lett. B 178 (1986) 324; S. Weinberg, Int. J. Mod. Phys. A 2 (1987) 301; C. Rubbia, CERN-PPE/93-08; G. Fiorentini et al., INFNFE-10-93.

[2] S.M. Bilenky and C. Giunti, Phys. Lett. B 311 (1993) 179.

[3] S.M. Bilenky and C. Giunti, DFTT 62/93, to be published in Phys. Lett. B.

[4] The Sudbury Neutrino Observatory Collaboration, Phys. Lett. B 194 (1987) 321; H.H. Chen, Nucl. Instr. Meth. A 264 (1988) 48; G.T. Ewan, Proceedings of the $4^{\text{th}}$ International Workshop on Neutrino Telescopes, Venezia, March 1992.

[5] S.M. Bilenky and B. Pontecorvo, Phys. Rep. 41 (1978) 225; S.M. Bilenky and S.T. Petcov, Rev. Mod. Phys. 59 (1987) 671.

[6] A. Cisneros, Astrophys. Space. Sci. 10 (1970) 87; M.B. Voloshin, M.I. Vysotsky and L.B. Okun, Zh. Eksp. Teor. Fiz. 91 (1986) 754 [ Sov. Phys. JETP 64 (1986) 446].

[7] C.S. Lim and W.J. Marciano, Phys. Rev. D 37 (1988) 1368; E.Kh. Akhmedov, Phys. Lett. B 213 (1988) 64.

[8] P.B. Pal, Int. J. Mod. Phys. A 7 (1992) 5387.

[9] R. Davis Jr., Intern. Symp. on Neutrino Astrophys., Takayama/Kamioka, Japan, 1992.

[10] E.Kh. Akhmedov, Sov. Phys. JETP 68 (1989) 690; C.S. Lim et al., Phys. Lett. B 243 (1990) 389; E.Kh. Akhmedov, Phys. Lett. B 255 (1991) 84; E.Kh. Akhmedov, S.T Petcov and A.Yu. Smirnov, Proceedings of the 5$^{th}$ International Workshop on Neutrino Telescopes, Venezia, March 1993; H. Nunokawa and H. Minakata, Phys. Lett. B 314 (1993) 371.

[11] M. Aglietta et al., Astropart. Phys. 1 (1992) 1.

[12] R. Barbieri et al., Phys. Lett. B 259 (1991) 119.

[13] H.H. Chen, Nucl. Instr. Meth. A 264 (1988) 48.

[14] A. Suzuki, Proc. of the Workshop on Elementary Particle Picture of the Universe, KEK, Japan, 1987; Y. Totsuka, ICRR-report-227-90-20 (1990); C.B. Bratton et al., "Proposal to Participate in the Super-Kamiokande Project", Dec. 1992.

[15] K. Kubodera and S. Nozawa, USC(NT)-93-6, nucl-th/9310014.

[16] J.N. Bahcall and R. Ulrich, Rev. Mod. Phys. 60 (1988) 297; J.N. Bahcall, Neutrino Physics and Astrophysics, Cambridge University Press, 1989.

[17] P. Langacker, Lectures presented at TASI-92, Boulder, June 1992.

[18] E.Kh. Akhmedov, A. Lanza and S.T Petcov, Phys. Lett. B 303 (1993) 85.

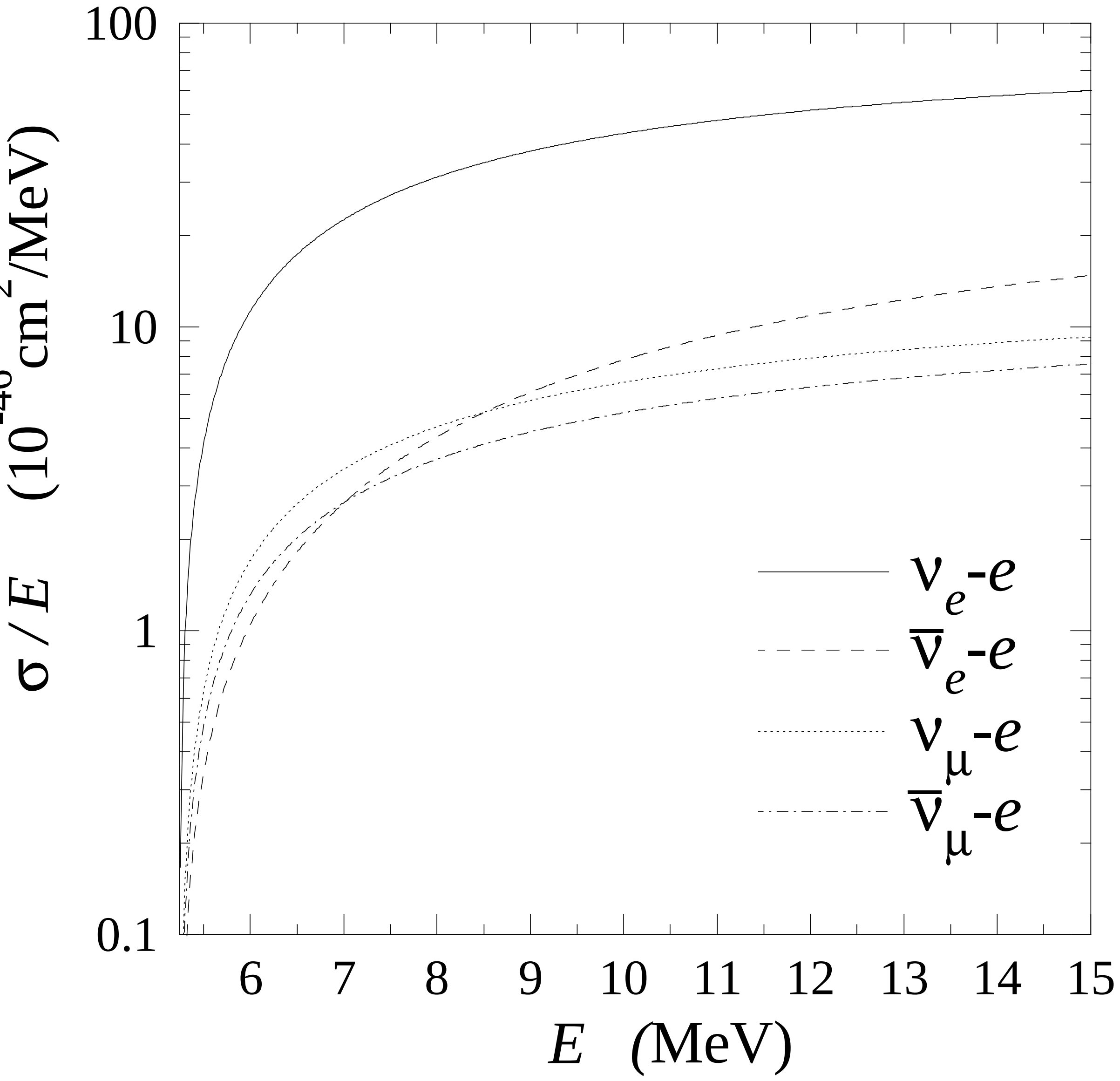

Figure 1: Values of the cross sections $\sigma_{\nu_e e}(E)$, $\sigma_{\nu_\mu e}(E)$, $\sigma_{\bar{\nu}_e e}(E)$, $\sigma_{\bar{\nu}_\mu e}(E)$ in the energy range relevant for solar neutrino experiments, with a 5 MeV threshold kinetic energy for electron detection. We used $g_V = -0.036$ and $g_A = -0.505$ [17].